\documentclass[aps,pra,superscriptaddress,groupaddress,showpacs,notitlepage,nofootinbib]{revtex4-1}
\usepackage{graphics,graphicx,color,times,amsfonts,amssymb,amsmath,bm,bbm,dsfont,hyperref,xcolor}
\newtheorem{theorem}{Theorem}

\newtheorem{lemma}{Lemma}

\newcommand{\ignore}[1]{}

\hypersetup{ 
colorlinks, 
linkcolor={blue}, 
citecolor={blue}, 
urlcolor= {blue} 
}

\let\oldsqrt\sqrt
\def\sqrt{\mathpalette\DHLhksqrt}
\def\DHLhksqrt#1#2{%
\setbox0=\hbox{$#1\oldsqrt{#2\,}$}\dimen0=\ht0
\advance\dimen0-0.2\ht0
\setbox2=\hbox{\vrule height\ht0 depth -\dimen0}%
{\box0\lower0.4pt\box2}}

\DeclareFontFamily{OT1}{pzc}{}
\DeclareFontShape{OT1}{pzc}{m}{it}%
              {<-> s * [1.25] pzcmi7t}{}
\DeclareMathAlphabet{\mathpzc}{OT1}{pzc}%
                                 {m}{it}

\begin{document}

\title{Quantum ergodicity for a class of non-generic systems}

\author{ P. Asadi}
\affiliation{Department of Physics, Sharif University of Technology, Tehran 14588, Iran}
\affiliation{Department of Physics and Astronomy, Rutgers University, Piscataway, NJ 08854, USA}

\author{F. Bakhshinezhad}
\affiliation{Department of Physics, Sharif University of Technology, Tehran 14588, Iran}

\author{ A. T. Rezakhani}
\affiliation{Department of Physics, Sharif University of Technology, Tehran 14588, Iran}

\begin{abstract}
We examine quantum normal typicality and ergodicity properties for quantum systems whose dynamics are generated by Hamiltonians which have residual degeneracy in their spectrum and resonance in their energy gaps. Such systems can be considered atypical in the sense that degeneracy, which is usually a sign of symmetry, is naturally broken in typical systems due to stochastic perturbations. In particular, we prove a version of von Neumann's quantum ergodic theorem, where a modified condition needs to hold in order to have normal typicality and ergodicity. As a result, we show that degeneracy of spectrum does not considerably modify the condition of the theorem, whereas the existence of resonance is more dominant for obstructing ergodicity.
\end{abstract}

\maketitle

\section{Introduction}
\label{sec:int}

Statistical mechanics has proved to be a successful theory for macroscopic systems. One of the cornerstones of statistical mechanics is the ergodic hypothesis, which colloquially states that the fraction of time state of a systems spends in a given subspace of its state space is proportional to the fraction of the surface occupied by this subspace \cite{book:Toda,book:Jancel,book:Reichl,book:Benatti}. The situation in quantum statistical mechanics becomes specially interesting in light of basic differences with classical statistical mechanics. Such differences are caused by the very mathematical structure of quantum mechanics, which allow ``typicality" behaviors to emerge \cite{Tasaki,G-typicality,ref-Popescu Nature,Reimann,ref-Bartsch,ref-Cho PRL}. 

The first, seminal attempt to put the ergodic hypothesis in the form of a rigorous theorem was made by von Neumann through proving a ``quantum ergodic theorem" \cite{ref-von Neumann Proof,ref-von Neumann}. The validity of this theorem, however, was heavily debated later in the literature \cite{Farquhar-Landsberg,ref-Boc PR, book:Farquhar}. It was just recently that this theorem was revisited carefully in Ref.~\cite{Goldstein-et-al} where the earlier criticisms were refuted, and it was rigorously proved that the theorem (with the set of sufficient conditions assumed by von Neumann) is indeed valid. Specifically, it has been shown that von Neumann's statement is in fact a more general property than the ergodicity, called \textit{normal typicality}. This reaffirmation of the quantum ergodic theorem  resolves a long-standing issue with quantum statistical mechanics, and is of fundamental importance. Interest in studying underlying laws of thermodynamics and statistical mechanics has been recently reinvigorated (mostly by advances in quantum information science), which also includes revisiting (emergent) properties such as ergodicity, thermal equilibration, or out-of-equilibrium fluctuations \cite{Brody,ref-Brandao Nature,ref-Popescu PRE,ref-Short T,Short-2,Goldstein-time,Goldstein-time-2,Faraj-dynamics,ref-Gessner,ref-Kosloff,ref-GABRIEL,Horodecki,ref-Galkowski,gogolin,new,Esposito,Goldstein-et-al2,Campisi}. Thus a clear form of the quantum ergodic theorem can enrich such attempts to better understand the principles of statistical mechanics in light of quantum mechanics.

Here we further the proof of Ref.~\cite{Goldstein-et-al}, and lift some necessary conditions of the quantum ergodic theorem, and investigate how this may affect the theorem. Non-degeneracy of energy spectrum and energy gaps both have been assumed in the proof of the quantum ergodic theorem. Non-degeneracy and non-resonance are mainly due to small and uncontrollable interactions of quantum systems with their environment, and thus exist in typical systems. Here we, however, consider \textit{atypical} quantum systems whose dynamics are generated by Hamiltonians which have residual degeneracy in their spectrum (not completely lifted by external perturbations) and resonance in their energy gaps. We find conditions under which normal typicality and ergodicity may still hold to some extent.

The structure of this paper is as follows. In Sec.~\ref{sec:notation}, we set notations and recall necessary definitions and results from literature. In Sec.~\ref{sec:results}, we state our main result in Theorem \ref{thm:thm1}, which shows how non-degeneracy and non-resonance conditions modify the condition for normal typicality and ergodicity. We next provide the proof of Theorem \ref{thm:thm1} in Sec.~\ref{sec:proof}. The paper ends by a discussion and summary of our result, followed by two appendices where some necessary details are proved.  

\section{Preliminaries}
\label{sec:notation}

In this section, we set up some preliminaries and remind pertinent definitions and results (see Ref.~\cite{Goldstein-et-al} for details).  

\subsection{Hamiltonian and dynamics}

Let us assume that we have a quantum system whose associated Hilbert space is $\mathpzc{H}$ (where $D=\dim(\mathpzc{H})$). The dynamics of this system is generated by the Hamiltonian
\begin{align}
H=\sum_{\alpha=1}^{D_{\mathrm{E}}\leqslant D}E_{\alpha} \Pi_{\alpha},
\end{align}
where $E_{\alpha}$ is an $e_{\alpha}$-fold degenerate energy eigenvalue ($E_{\alpha}\neq E_{\beta}$ iff $\alpha\neq \beta$) and $\Pi_{\alpha}= \sum_{a=1}^{e_{\alpha}} |\alpha, a\rangle \langle \alpha, a|$ represents the corresponding eigenprojection. Here $\sum_{\alpha=1}^{D_{\mathrm{E}}}e_{\alpha}=D$, in which $D_{\mathrm{E}}$ is the number of distinct eigenvalues. Any initial state $|\psi(0)\rangle$ evolves into
\begin{equation}
|\psi(\tau)\rangle=e^{-i\tau H}|\psi(0)\rangle,
\end{equation}
after time $\tau$, where we have assumed $\hbar\equiv1$ (here and hereafter). 

We also need to define (unnormalized) vectors $|\widetilde{\varphi}_{\alpha}\rangle$ and (normalized) vectors $|\varphi_{\alpha}\rangle$ belonging to the $\alpha$th energy shell as
\begin{align}
\Pi_{\alpha} |\psi(0)\rangle & = \sum_{a=1}^{e_{\alpha}}s_{\alpha a}|\alpha,a\rangle\nonumber\\
&\equiv | \widetilde{\varphi}_{\alpha}\rangle \equiv c_{\alpha} | \varphi_{\alpha} \rangle,
\label{not3}
\end{align}
where $s_{\alpha a}=\langle \alpha,a|\psi(0)\rangle$ and $c_{\alpha}$ is the normalization factor ($|c_{\alpha}|^2=\langle\widetilde{\varphi}_{\alpha}|\widetilde{\varphi}_{\alpha}\rangle$).

\subsection{Gap and energy-sum structures}
\label{subsec:gap}

\begin{figure}[tp]
\includegraphics[scale=.5]{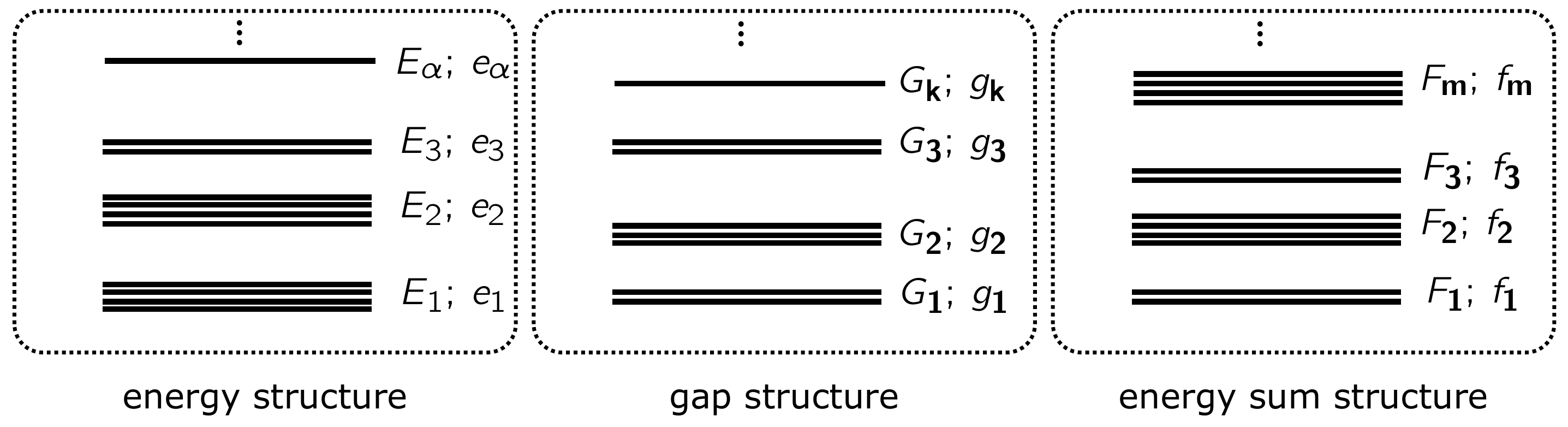}
\caption{Schematic of the energy, gap, and energy sum structures. Each energy eigenvalue $E_{\alpha}$ has degeneracy $e_{\alpha}$ ($\alpha\in\{1,2,\ldots,D_{\mathrm{E}}\}$); each gap value $G_{\mathbf{k}}\equiv E_{\alpha}-E_{\gamma}$ (for some $\mathbf{k}(\alpha,\gamma)$), where $\alpha\neq \gamma$, has degeneracy $g_{\mathbf{k}}$; and each energy sum value $F_{\mathbf{m}}\equiv E_{\alpha} + E_{\gamma}$ (for some $\mathbf{m}(\alpha,\gamma)$) has degeneracy $f_{\mathbf{m}}$. Here $\mathpzc{G}_{\mathbf{0}}$ denotes the vanishing-gap set, which has $g_{\mathbf{0}}=D_{\mathrm{E}}$ elements. In addition, we have $\sum_{\mathbf{k}\neq \mathbf{0}}g_{\mathbf{k}}=D_{\mathrm{E}}(D_{\mathrm{E}}-1)$. We also define $D_{\mathrm{G}}\equiv \max_{\mathbf{k}\neq 0} g_{\mathbf{k}}$ and $D_{\mathrm{F}}\equiv \max_{\mathbf{m}} f_{\mathbf{m}}$. The case of the non-resonant Hamiltonian is given by $D_{\mathrm{G}}=1$, or equivalently, $D_{\mathrm{F}}=2$ (see Appendix \ref{app:df}). }
\label{fig:gap}
\end{figure}

We set the shorthand $E_{\beta} - E_{\alpha}=: G_{\mathbf{k}} $ to denote the energy gap, for some $\mathbf{k}$. For a given value of $G_{\mathbf{k}}$, there may exist several ordered pairs $(\alpha,\beta)$ for which $E_{\beta} - E_{\alpha}$ equals this given value of $G$. Thus we define $\mathpzc{G}_{\mathbf{k}}:=\{(\alpha,\beta)~|~E_{\beta}-E_{\alpha}=G_{\mathbf{k}}\}$. It is evident that $|\mathpzc{G}_{\mathbf{k}}|=g_{\mathbf{k}}$, where $g_{\mathbf{k}}$ is the degeneracy of the energy gap $G_{\mathbf{k}}$, and $|A|$ denotes cardinality of set $A$. Similarly, for sum $F_{\mathbf{m}}$ of distinct energies, we define $\mathpzc{F}_{\mathbf{m}}:=\{(\alpha,\gamma)~|~E_{\alpha}+E_{\gamma}=F_{\mathbf{m}}\}$ and $f_{\mathrm{m}}=|\mathpzc{F}_{\mathbf{m}}|$. For later use, we also define $D_{\mathrm{G}}\equiv \max_{\mathbf{k}\neq 0} g_{\mathbf{k}}$ and $D_{\mathrm{F}}\equiv \max_{\mathbf{m}} f_{\mathbf{m}}$. Figure~\ref{fig:gap} illustrates a schematic of the gap and energy-sum structures. 

A Hamiltonian is called \textit{non-degenerate} if none of its energy eigenvalues $E_{\alpha}$ is degenerate, i.e., $e_{\alpha}=1~\forall \alpha$, or equivalently $D_{\mathrm{E}}=D$. Similarly, a Hamiltonian is called \textit{non-resonant} if none of its energy gaps is degenerate, i.e., $g_{\mathbf{k}}=1~\forall \mathbf{k}\neq\mathbf{0}$, or equivalently $D_{\mathrm{G}}=1$. Later in Appendix \ref{app:df} we show that the non-resonance condition implies $f_{\mathbf{m}}=2~\forall \mathbf{m}$, or equivalently $D_{\mathrm{F}}=2$. Thus, in general we have $D_{\mathrm{F}}\geqslant 2$. We were, however, not able to find a general relation between the gap-related quantity $D_{\mathrm{G}}$ and the corresponding energy-sum-related quantity $D_{\mathrm{F}}$. But intuitively it seems that a relatively small $D_{\mathrm{G}}$ corresponds to a relatively small $D_{\mathrm{F}}$. 

\subsection{Measurements, macrostates, and microstates}
\label{subsec:measurements}

We partition $\mathpzc{H}$ according to a complete set of given (but otherwise arbitrary) orthogonal projections $\{P_{\nu}\}_{\nu=1}^{M}$, in which $P_{\nu}$ has rank $d_{\nu}$ and $\sum_{\nu=1}^{M} P_{\nu}=\openone$ \cite{ref-von Neumann Proof}, as
\begin{align}
\mathpzc{H}=\oplus_{\nu=1}^{M}\mathpzc{H}_{\nu},
\label{partition}
\end{align}
with $\mathpzc{H}_{\nu}=\{P_{\nu}|v\rangle,~|v\rangle\in\mathpzc{H}\}$. It is evident that
\begin{align}
\dim(\mathpzc{H}_{\nu}) &=d_{\nu},\\
\sum_{\nu=1}^{M}d_{\nu} &=D. \label{dnu}
\end{align}

\begin{figure}[tp]
\includegraphics[scale=.3]{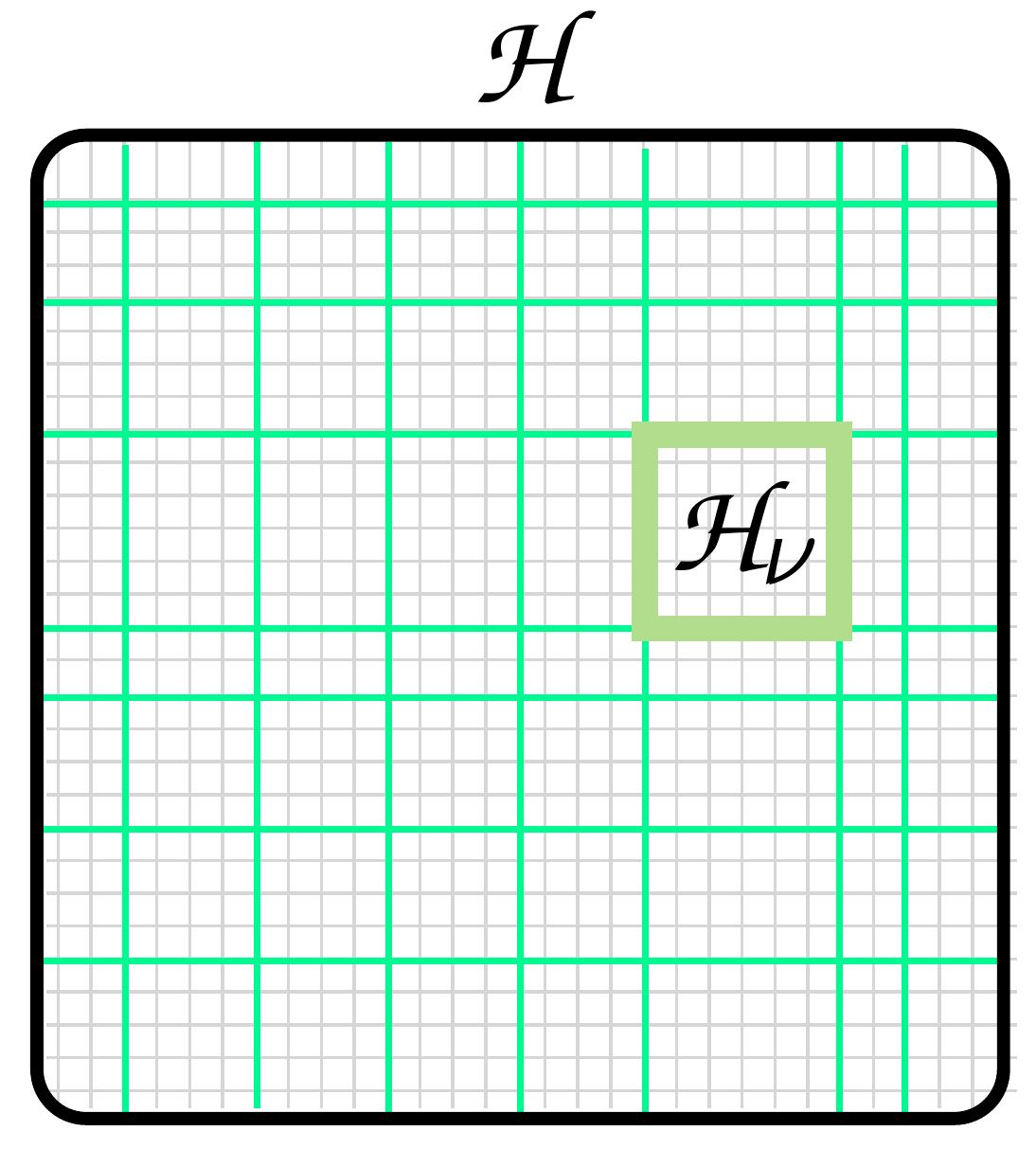}
\caption{Schematic structure of the Hilbert space $\mathpzc{H}=\oplus_{\nu=1}^{M}\mathpzc{H}_{\nu}$, decomposed to $M$ ``macrospaces" or ``Gibbs cells" (green cells) according to a complete set of macro-observables $\{P_{\nu}\}_{\nu=1}^{M}$. Each (experimentally accessible) Gibbs cell includes many ``microstates" (gray cells), as implied by the assumption $\dim(\mathpzc{H}_{\nu})=d_{\nu}\ll \dim(\mathpzc{H})=D$, $\forall \nu$.}
\label{fig:hilbertspace}
\end{figure}

 The set $\{P_{\nu}\}_{\nu=1}^{M}$ can be associated with a complete measurement on the system \cite{book:Nielsen}; or equivalently, each $\mathpzc{H}_\nu$ can be interpreted to represent a ``macrospace/macrostate" or ``Gibbs cell" of the system in which all ``microstates" yield the same result for the measurement---see Fig.~\ref{fig:hilbertspace}. 

\subsection{Averaging and von Neumann's lemmas}
\label{subsec:vonneumann}

Let $\mathbb{E}_{X}[f(x)]=\int_{\mathpzc{D}} f(x)~p_X(x)\mathrm{d}x$ denote the average or expected value of $f(x)$ when $X$ is a random variable defined on a set $\mathpzc{D}\subseteq \mathds{R}$ with a given probability distribution $p_X(x)$. In addition, the variance of $X$ is given by $\mathbb{V}_X[x]=\mathbb{E}_{X}\big[(x-\mathbb{E}_{X}[x])^2\big]$ \cite{book:Reichl}. From positivity of the variance, $\mathbb{V}_{X}[x]\geqslant 0$, we immediately have
\begin{equation}
\mathbb{E}^{2}_{X}[x]\leqslant\mathbb{E}_{X}[x^2].
\label{eq:var>0}
\end{equation}
Similarly, from $\mathbb{E}_{X} [(x_1-x_2)^2] \geqslant 0$ for two random variables $x_1$ and $x_2$, we have 
\begin{align}
2\mathbb{E}_{X}[x_1 x_2] \leqslant \mathbb{E}_{X}[x_1^2] + \mathbb{E}_{X}[x_2^2].
\label{g4b2}
\end{align}
Hence, when $x_1$ and $x_2$ have identical expected values and variances, we have
\begin{equation}
\mathbb{E}_{X}[x_1 x_2] \leqslant \mathbb{E}_{X}^2[x_1] + \mathbb{V}_{X}[x_1].
\label{eq:e-v}
\end{equation}
One can also see from the definition of the variance that it satisfies the following property:
\begin{align}
\mathbb{V}_{X}\Big[\sum_{i=1}^{K} x_{i}\Big] = \sum_{i=1}^{K}\mathbb{V}_{X}[x_{i}] +\sum_{i\neq j=1}^{K} \mathbb{C}_{X}[x_{i},x_{j}],
\label{var-covar}
\end{align}
where $\mathbb{C}_{X}[x_{i},x_{j}]\equiv \mathbb{E}_{X}[x_i]\,\mathbb{E}_{X}[x_j]-\mathbb{E}_{X}[x_i x_j]$ is the covariance of the random variables $x_i$ and $x_j$ \cite{book:Jancel,book:Reichl}.

In addition to the basic definitions above, we shall also need a result known as Markov's inequality \cite{book:Reichl},
\begin{equation}
\mathrm{Prob}[X\geqslant B]\leqslant \mathbb{E}_{X}[x]/B,~~B>0.
\label{eq:markov}
\end{equation}
From this we say $X\leqslant B$ holds for $(1-\delta)$ most $X$s if $\mathrm{Prob}[X\leqslant B]\geqslant 1-\delta$ \cite{Goldstein-et-al}.

The following lemmas---proven by von Neumann (adopted from Ref.~\cite{Goldstein-et-al})---will also be essential later:

\begin{lemma}
There exists a number $C>1$ such that when two natural numbers $d$ and $D$ satisfy the condition
\begin{equation}
C\frac{\log D}{D}< \frac{d}{D}<\frac{1}{C},
\label{condition-c}
\end{equation}
for uniformly-randomly distributed unitaries $U\in\mathrm{SU}(D)$ (according to the Haar measure \cite{book:Jancel,ref-Muller}) we have
\begin{align}
&\mathbb{E}_{U}\Big[ \max_{i\neq j\in\{1,\ldots,D\}}\Big|\sum_{\ell=1}^{d}U_{\ell i}U^{*}_{\ell j}\Big|^2 \Big]\leqslant \frac{\log D}{D}<\frac{1}{C^2},\label{eq:l1-1}\\
&\mathbb{E}_{U}\Big[\max_{i\in\{1,\ldots,D\}}\Big(\sum_{\ell=1}^{d}|U_{\ell i}|^2-\frac{d}{D}\Big)^2 \Big]\leqslant \frac{9d\log D}{D^2}< \frac{9}{C^3}.
\label{eq:l1-2}
\end{align}
\label{lemma:vonneumann1}
\end{lemma}

\begin{lemma}
Let $|\xi\rangle$ be a uniformly-randomly distributed state from a $D$-dimensional Hilbert space and $P$ be a rank-$d$ projection thereon. Then
\begin{align}
\mathbb{E}_{\xi}[\Vert P|\xi\rangle\Vert^2] & = \frac{d}{D},\\
\mathbb{V}_{\xi}[\Vert P|\xi\rangle\Vert^2] &=\frac{1}{d}\left(\frac{d}{D}\right)^2 \frac{(D-d)}{(D+1)}.
\end{align}
\label{lemma:vonneumann2}
\end{lemma}
For the sketch of proof see Appendix \ref{app:averages} \cite{book:Jancel}.
 
\subsection{Long-time average and long run}
\label{subsec:longrun}

For a dynamical function $Y(\tau)$, we define the long-time average $\overline{Y}$ as follows \cite{ref-Popescu PRE}:
 \begin{equation}
\overline{Y}=\lim_{T\to \infty} \frac{1}{T} \int_{0}^{\infty} Y(\tau)\mathrm{d}\tau.
\label{not2}
\end{equation}
We also say that a statement $S(\tau)$ \textit{holds for $(1-\delta')$ fraction of the time in the long run} when
\begin{equation}
\liminf_{T\to \infty}\frac{1}{T}\big|\big\{\tau\in[0,T];~ S(\tau)~ \mathrm{holds}\big\}\big|\geqslant (1-\delta'),
\label{long run}
\end{equation}
where $|A|$ indicates the Lebesgue measure or size of set $A\subset \mathds{R}$ \cite{Goldstein-et-al}.

\textit{Remark.---}In the standard context of the quantum ergodic theorem we focus on the $T\to\infty$ time averages. It is, however, interesting to see how the setting of the theorem and associated results may change for \textit{finite}-time averages. This subject is beyond the scope of the current paper. See, for example, Refs.~\cite{Goldstein-time,Goldstein-time-2,Short-2,Faraj-dynamics,ref-Short T} for discussions on finite time scales for equilibration.

\subsection{Normal typicality}
\label{subsec:normality}

The normal typicality property in the sense of von Neumann implies that for \textit{almost all} partitionings of a Hilbert space $\mathpzc{H}$ according to a complete set of $M$ rank-$d_{\nu}$ orthogonal projections (say $\{P_{\nu}\}_{\nu=1}^{M}$)  as in Eq.~(\ref{partition}), and for \textit{all} initial states $|\psi(0)\rangle\in\mathpzc{H}$, a system is $\varepsilon$-$\delta'$ normal if for $(1-\delta')$ fraction of the time in the long run we have \cite{Goldstein-et-al}
\begin{equation}
\left|\|P_{\nu}|\psi(\tau)\rangle \|^2- d_{\nu}/D\right| \leqslant \frac{\varepsilon}{\sqrt{M}} \sqrt{\frac{d_{\nu}}{D}}, ~~~\forall \nu\in\{1,\ldots,M\},
\label{eq0}
\end{equation}
where $\Vert u\Vert\equiv \sqrt{\langle u|u \rangle}$ (for $|u\rangle\in\mathpzc{H}$) indicates the Euclidean vector norm.

A \textit{sufficient} condition for this result to hold is that
\begin{equation}
\mathpzc{L} \equiv \overline{\left|\|P_{\nu}|\psi(\tau)\rangle \|^2- d_{\nu}/D\right|^2} \leqslant \delta' \Big(\frac{\varepsilon}{M}\Big)^{2} \frac{d_{\nu}}{D},~~~\forall \nu\in\{1,\ldots,M\}
\label{eq00}
\end{equation}
in the long run \cite{Goldstein-et-al}. The proof is immediate by contradiction. To see this, assume $Y(\tau)$, $p$, and $q$ to be the left-hand side (LHS) of Eq.~(\ref{eq0}), $\varepsilon\sqrt{ d_{\nu}/(MD)}$, and $\delta' /M$, respectively. Let us assume that $Y(\tau)\geqslant p$ ($\geqslant0$) for at least $q$ fraction of the total time $T$; this assumption is clearly a violation of Eq.~(\ref{eq0}). Hence the time average of $Y^2(\tau)$ on the interval $[0,T]$ is $\geqslant qp^2$; which in turn violates Eq.~(\ref{eq00}). Thus Eq. (\ref{eq00}) is a sufficient condition for the property (\ref{eq0}).

\subsection{(Relative) Ergodicity}
\label{subsec:qerg}

We call a closed quantum system evolving via a time-independent Hamiltonian $\varepsilon$-ergodic relative to a \textit{given} set of complete observables (measurements) $\{P_{\nu} \}_{\nu=1}^{M}$ if for every initial state $|\psi(0)\rangle$ we have
\begin{equation}
\left|\overline{\|P_{\nu}|\psi(\tau)\rangle \|^2}- d_{\nu}/D\right|^2 \leqslant \Big(\frac{\varepsilon}{M}\Big)^2 \frac{d_{\nu}}{D},~~~~\forall\nu\in\{1,\ldots,M\}.
\label{result1}
\end{equation}
In other words, when the RHS of Eq.~(\ref{result1}) becomes negligible, we conclude that the average time the system spends in a certain macrostate (induced by the $P_{\nu}$ measurements) is proportional to the relative \textit{size} of that Gibbs cell (here denoted by $d_{\nu}$) compared to the size of the entire available Hilbert space (here denoted by $D$). If the above property holds for (almost) all possible measurements, we call the system ergodic. This statement somehow resembles the traditional statement of the ergodic hypothesis in classical statistical mechanics \cite{book:Reichl}.

There is a clear connection between normal typicality and ergodicity. We note that  
\begin{align} 
\big| \overline{\|P_{\nu}|\psi(\tau)\rangle \|^2}-d_{\nu}/D\big|^2 & =  \big|\overline{\|P_{\nu}|\psi(\tau)\rangle \|^2- d_{\nu}/D}\big|^2\nonumber\\
&=  \lim_{T\to \infty}\frac{1}{T^2}\left| \int_0^T \left(\|P_{\nu}|\psi(\tau)\rangle \|^2-d_{\nu}/D\right)\mathrm{d}\tau\right|^2 \nonumber\\
&\overset{\mathrm{Eq.~(\ref{eq:var>0});}~\overline{X}^2\leqslant\overline{X^2}}{\leqslant} \lim_{T\to \infty}\frac{1}{T}\int_0^T \left| \|P_{\nu}|\psi(\tau)\rangle \|^2- d_{\nu}/D\right|^2 \mathrm{d}\tau\nonumber\\
&=\overline{\left| \|P_{\nu}|\psi(\tau)\rangle \|^2-d_{\nu}/D\right|^2},
\label{A1}
\end{align}
or equivalently,
\begin{equation}
\big|\overline{\|P_{\nu}|\psi(\tau)\rangle \|^2}- d_{\nu}/D\big|^2 \leqslant \mathpzc{L}.
\label{QE}
\end{equation}
Hence normal typicality (i.e., a relatively small upper bound on $\mathpzc{L}$) implies ergodicity \cite{Goldstein-et-al}. We shall use this fact later.

\section{Main result}
\label{sec:results}

Our aim is to see whether and how the results of Refs.~\cite{ref-von Neumann,Goldstein-et-al} are modified in the case of degenerate and resonant Hamiltonians. The following theorem, a modified version of von Neumann's quantum ergodic theorem, encapsulates our main result. Note that a stronger (or perhaps tighter) version of the theorem may still be possible, but we shall not discuss it here.

\begin{theorem} 
\label{thm:thm1}
Let $0<\delta,\delta'\leqslant 1$ and $\varepsilon>0$, and consider a $D$-dimensional Hilbert space $\mathpzc{H}$ associated with a quantum system. This system is $\varepsilon$-$\delta'$ normal for (1-$\delta$) fraction of all Hilbert-space decompositions $\mathpzc{H}=\oplus_{\nu=1}^{M}\mathpzc{H}_{\nu}$, with given (fixed) dimensions $\{d_{\nu}\}$ induced by measurements, if
\label{theorm1}
\begin{align}
\max \left\{C , \frac{10M^2}{\delta\delta'\varepsilon^2} \left(1 +    \frac{(D_{\mathrm{F}}-2)d_\nu^2}{10D\log D} \right) \right\}\frac{\log D}{D}<   \frac{d_{\nu}}{D} < \frac{1}{C}, ~~~\forall \nu\in\{1,\ldots,M\},
\label{thrm1}
\end{align}
where $C$ is a number satisfying Lemma \ref{lemma:vonneumann1}, and $D_{\mathrm{F}}$ is the maximum degeneracy of the energy sum structure of the system (defined in Subsec.~\ref{subsec:gap}).
\end{theorem}

In the absence of resonance (the case where $D_{\mathrm{F}}=2$), Eq.~(\ref{thrm1}) implies that as long as the dimension of each Gibbs cell is sufficiently large ($d_{\nu}\gg1$), yet sufficiently small compared to the dimension of the entire Hilbert space ($d_{\nu}\ll D$), the system will feature normal typicality for a certain fraction of the measurements. This is the standard quantum ergodic theorem \cite{Goldstein-et-al}. Note, however, that in Eq.~(\ref{thrm1}) degeneracy does not show up explicitly, whereas resonance appears to be important for obstructing or allowing normal typicality (and ergodicity)---because of the existence of the factor $(D_{\mathrm{F}}-2)d_\nu^2/(10D\log D)$. From this one can conclude that the impact of resonance would be relatively small if
\begin{equation}
D_{\mathrm{F}}\ll (10 \log D/D)M^2,
\label{dim-condition}
\end{equation}
where we have used the crude estimate $d_{\nu}\approx O(D/M)$ [see Eq.~(\ref{dnu})]. 

We also remark that for a given pair ($d_{\nu},D$) there may exist several $C$s satisfying Lemma \ref{lemma:vonneumann1}. In this case we choose the largest one. As an example, for a system of $10^2$ spin-$1/2$ particle with $D=2^{10^{2}}\approx 10^{30}$, when $d_{\nu}\approx 10^8$, and $M\approx 10^{22}$, we have $\log D/D\approx 10^{-29}$, $d_{\nu}/D\approx 10^{-22}$. Thus, we can choose $C\approx 10^{7}$. Additionally, one may be able to choose $\varepsilon$ and $\delta'$ such that they satisfy the condition (\ref{thrm1}), for sufficiently negligible values of $\delta$, and hence observe ergodicity according to Eqs.~(\ref{eq00}) and (\ref{result1}).

It would be instructive to consider specific physical examples (e.g., spin chains with strong symmetries and too large degeneracies and resonance) in which one or parts of the conditions of Theorem \ref{thm:thm1} do not hold, whereas one could explicitly demonstrate that the system is not ergodic. We, however, leave this (important) task for future investigations, and here focus mainly on the formalism of the quantum ergodic theorem. 

\section{Proof of Theorem \ref{thm:thm1}}
\label{sec:proof}

The logic of the proof goes as follows. We find an upper bound $X$ on $\mathpzc{L}$. In order for normal typicality (and whence---according to Eq.~(\ref{QE})---ergodicity) to hold, we shall require that this upperbound $X$ itself be not greater than the value given by the sufficient condition for $\varepsilon$-$\delta'$ normal typicality (\ref{eq00}), specifically, the RHS of this equation: $B\equiv \delta'(\varepsilon/M)^2(d_{\nu}/D)$. That is,
\begin{align}
\mathrm{ergodicity~[LHS~of~Eq.~(\ref{result1})]}\overset{\mathrm{Eq.~(\ref{QE})}}{\leqslant} \mathpzc{L} \leqslant X\leqslant B~\mathrm{[RHS~of~Eq.~(\ref{eq00})]}.
\label{eq.}
\end{align}
Next we shall employ Markov's inequality and require that its RHS be in turn upperbounded by $\delta$,
\begin{align}
\mathrm{Prob}[X> B] \leqslant \frac{\mathbb{E}[X]}{B}<\delta,
\label{MM}
\end{align}
which is tantamount to 
\begin{align}
\mathrm{Prob}[X\leqslant B]\overset{\mathrm{Eq.~(\ref{eq.})}}{=}\mathrm{Prob}[~\varepsilon\text{-}\delta'~\mathrm{normal~typicality~}]=\mathrm{Prob}[\mathrm{~ergodicity~}]\geqslant 1-\delta.
\end{align}
This implies that $\varepsilon$-$\delta'$ normal typicality and ergodicity hold for $(1-\delta)$ most measurements $\{P_{\nu}\}$ with fixed ranks $\{d_{\nu}\}$. We note that in order for this construction to work, we need to consider how our parameter $d_{\nu}$ should be appropriately chosen. There are two relevant conditions; the first one comes from the condition on $d_{\nu}$ to apply Lemma \ref{lemma:vonneumann1}, and the second condition comes from the requirement in the last part of Eq.~(\ref{MM}). We combine these conditions to find how/when an $\varepsilon$-$\delta'$ (with given values) normality/ergodicity is implied. In the sequel, we follow the steps elaborated here.  

Expanding $\mathpzc{L}$ [Eq.~(\ref{eq00})] in terms of $\{|\widetilde{\varphi}_{\alpha}\rangle\}$ yields
\begin{align}
\mathpzc{L} =&  (d_{\nu}/D)^2 - 2(d_{\nu}/D)~ \sum_{\alpha\beta} \overline{e^{-i\tau (E_{\beta}-E_{\alpha})}}~\langle\widetilde{\varphi}_{\alpha}|P_{\nu}|\widetilde{\varphi}_{\beta}\rangle 
+\sum_{\alpha\beta\gamma\sigma}\overline{e^{-i\tau (E_{\beta}-E_{\alpha}+E_{\gamma}-E_{\sigma})}} 
\langle\widetilde{\varphi}_{\alpha}|P_{\nu}|\widetilde{\varphi}_{\beta}\rangle \langle\widetilde{\varphi}_{\sigma}|P_{\nu}|\widetilde{\varphi}_{\gamma}\rangle.
\label{G}
\end{align}
In the following, we consider this expression term by term as 
\begin{equation}
\mathpzc{L} = (d_{\nu}/D)^{2} + \mathpzc{L}_{~2} + \mathpzc{L}_{~3}. 
\label{eq:L}
\end{equation}

The second term ($\mathpzc{L}_{~2}$) concerns degeneracy. If the Hamiltonian is non-degenerate, $\mathpzc{L}_{~2}$ can be simplified by using $\overline{e^{-i\tau (E_{\beta}-E_{\alpha})}}=\delta_{\alpha\beta}$ to
\begin{align}
\mathpzc{L}_{~2}^{(\mathrm{nd})}= -2(d_{\nu}/D) \sum_{\alpha=1}^{D}  \langle \widetilde{\varphi}_{\alpha}|P_{\nu}| \widetilde{\varphi}_{\alpha} \rangle.
\label{eq-deg4}
\end{align}
In the presence of degeneracy, we have
\begin{align}
\mathpzc{L}_{~2}^{(\mathrm{d})}&= -2(d_{\nu}/D) \sum_{\alpha \beta}\overline{e^{-i\tau (E_{\beta}-E_{\alpha})}}   \langle\widetilde{\varphi}_{\alpha}|P_{\nu}|\widetilde{\varphi}_{\beta} \rangle\nonumber\\ 
&\overset{\mathrm{Eq.~(\ref{not3})}}{=} -2(d_{\nu}/D) \sum_{\alpha \beta,ab} \delta_{E_{\alpha}E_{\beta}} s^{*}_{\alpha a}s_{\beta b}\langle \alpha,a|P_{\nu}| \beta,b\rangle\nonumber\\
&= -2(d_{\nu}/D) \sum_{\alpha,ab}  s^{*}_{\alpha a} s_{\alpha b}\langle \alpha,a|P_{\nu}| \alpha,b\rangle\nonumber\\
&= -2(d_{\nu}/D) \sum_{\alpha=1}^{D_{\mathrm{E}}}|c_{\alpha}|^2\langle\varphi_{\alpha}|P_{\nu}|\varphi_{\alpha} \rangle\nonumber\\
&= -2(d_{\nu}/D) \sum_{\alpha=1}^{D_{\mathrm{E}}}\langle\widetilde{\varphi}_{\alpha}|P_{\nu}|\widetilde{\varphi}_{\alpha} \rangle,
\label{degeneracy1}
\end{align}
which is akin to Eq.~(\ref{eq-deg4}) except in the appearance of $D_{\mathrm{E}}$ rather than $D$ in the upper limit of the summation.

The third term ($\mathpzc{L}_{~3}$) concerns resonance. For the non-resonance condition, we could replace \cite{Goldstein-et-al} 
\begin{equation}
\overline{e^{-i\tau (E_{\beta}-E_{\alpha}+E_{\gamma}-E_{\sigma})}}=\delta_{\alpha \beta}\delta_{\sigma \gamma}+\delta_{\sigma \beta}\delta_{\gamma \alpha}-\delta_{\gamma \beta \alpha \sigma},
\label{eq:delta}
\end{equation}
from whence $\mathpzc{L}_{~3}$ in Eq.~(\ref{G}) would reduce to
\begin{align}
\mathpzc{L}^{(\mathrm{nr})}_{~3}&=\sum_{\alpha}  |\langle \widetilde{\varphi}_{\alpha}|P_{\nu}|\widetilde{\varphi}_{\alpha} \rangle|^ 2 +\sum_{\alpha \neq \beta} |\langle \widetilde{\varphi}_{\alpha}|P_{\nu}|\widetilde{\varphi}_{\beta} \rangle|^2 + \sum_{\alpha \neq \beta}  \langle \widetilde{\varphi}_{\alpha}|P_{\nu}|\widetilde{\varphi}_{\alpha} \rangle \langle \widetilde{\varphi}_{\beta}|P_{\nu}|\widetilde{\varphi}_{\beta} \rangle
\nonumber\\
&=\sum_{\alpha \neq \beta} |\langle \widetilde{\varphi}_{\alpha}|P_{\nu}|\widetilde{\varphi}_{\beta} \rangle|^2 + \sum_{\alpha \beta}  \langle \widetilde{\varphi}_{\alpha}|P_{\nu}|\widetilde{\varphi}_{\alpha} \rangle \langle \widetilde{\varphi}_{\beta}|P_{\nu}|\widetilde{\varphi}_{\beta} \rangle.
\label{res-reduced-2}
\end{align}
In the presence of resonance, however, other terms will also appear,
\begin{align}
\mathpzc{L}_{~3} &=\sum_{\alpha\beta\gamma\sigma} \overline{e^{-i\tau (E_{\beta}-E_{\alpha}+E_{\gamma}-E_{\sigma})}} \langle\widetilde{\varphi}_{\alpha}|P_{\nu}|\widetilde{\varphi}_{\beta}\rangle \langle\widetilde{\varphi}_{\sigma}|P_{\nu}|\widetilde{\varphi}_{\gamma}\rangle\nonumber\\
 &=\Big(\sum_{\mathbf{k}:~(\alpha,\beta)\in \mathpzc{G}_{\mathbf{k}}}\sum_{\mathbf{l}:~(\gamma, \sigma)\in \mathpzc{G}_{\mathbf{l}}}\overline{e^{-i\tau ( G_{\mathbf{k}}- G_{\mathbf{l}})}} +\sum_{\mathbf{m}:~(\beta,\sigma)\in \mathpzc{G}_{\mathbf{m}}}\sum_{\mathbf{n}:~(\alpha,\gamma)\in \mathpzc{G}_{\mathbf{n}}}\overline{e^{-i\tau ( G_{\mathbf{m}}- G_{\mathbf{n}})}}\Big) \langle\widetilde{\varphi}_{\alpha}|P_{\nu}|\widetilde{\varphi}_{\beta}\rangle \langle\widetilde{\varphi}_{\sigma}|P_{\nu}|\widetilde{\varphi}_{\gamma}\rangle -\sum_{\alpha =1}^{D_{\mathrm{E}}} |\langle \widetilde{\varphi}_{\alpha}|P_{\nu}|\widetilde{\varphi}_{\alpha}\rangle|^2 \nonumber\\
&=\Big(\sum_{\mathbf{k}:~(\alpha,\beta)\in \mathpzc{G}_{\mathbf{k}}}\sum_{\mathbf{l}:~(\gamma, \sigma)\in \mathpzc{G}_{\mathbf{l}}} \delta_{G_{\mathbf{k}} G_{\mathbf{l}}}+\sum_{\mathbf{m}:~(\beta,\sigma)\in \mathpzc{G}_{\mathbf{m}}}\sum_{\mathbf{n}:~(\alpha,\gamma)\in \mathpzc{G}_{\mathbf{n}}}\delta_{G_{\mathbf{m}} G_{\mathbf{n}}}\Big) \langle\widetilde{\varphi}_{\alpha}|P_{\nu}|\widetilde{\varphi}_{\beta}\rangle \langle\widetilde{\varphi}_{\sigma}|P_{\nu}|\widetilde{\varphi}_{\gamma}\rangle-\sum_{\alpha =1}^{D_{\mathrm{E}}} |\langle \widetilde{\varphi}_{\alpha}|P_{\nu}|\widetilde{\varphi}_{\alpha}\rangle|^2 
\label{line3}\\
& = \sum_{\mathbf{k}} \Big( \sum_{(\alpha,\beta)\in\mathpzc{G}_{\mathbf{k}}} \sum_{(\gamma, \sigma)\in\mathpzc{G}_{\mathbf{k}}}  + \sum_{(\beta,\sigma)\in\mathpzc{G}_{\mathbf{k}}} \sum_{(\alpha,\gamma)\in\mathpzc{G}_{\mathbf{k}}}\Big) \langle\widetilde{\varphi}_{\alpha}|P_{\nu}|\widetilde{\varphi}_{\beta}\rangle \langle\widetilde{\varphi}_{\sigma}|P_{\nu}|\widetilde{\varphi}_{\gamma}\rangle -\sum_{\alpha =1}^{D_{\mathrm{E}}} |\langle \widetilde{\varphi}_{\alpha}|P_{\nu}|\widetilde{\varphi}_{\alpha}\rangle|^2 \nonumber\\
& =\mathpzc{L}^{(\mathrm{nr})}_{~3} + \sum_{\mathbf{k}}\Big( \sum_{(\alpha,\beta)\in \mathpzc{G}_{\mathbf{k}}} \sum_{(\gamma, \sigma)\in \mathpzc{G}_{\mathbf{k}}|(\alpha,\beta)\neq (\gamma, \sigma)} + \sum_{(\beta,\sigma)\in \mathpzc{G}_{\mathbf{k}}} \sum_{(\alpha,\gamma)\in \mathpzc{G}_{\mathbf{k}}|(\beta,\sigma)\neq (\alpha,\gamma)}  \Big) \langle\widetilde{\varphi}_{\alpha}|P_{\nu}|\widetilde{\varphi}_{\beta}\rangle \langle\widetilde{\varphi}_{\sigma}|P_{\nu}|\widetilde{\varphi}_{\gamma}\rangle\nonumber\\
&=\mathpzc{L}^{(\mathrm{nr})}_{~3} + \sum_{\mathbf{m}} \sum_{(\alpha,\sigma)\in \mathpzc{F}_{\mathbf{m}}} \sum_{(\beta,\gamma)\in \mathpzc{F}_{\mathbf{m}}\big|\begin{smallmatrix}(\beta,\gamma)\neq(\alpha,\sigma) \\(\beta,\gamma)\neq (\sigma,\alpha)\end{smallmatrix}}  \langle\widetilde{\varphi}_{\alpha}|P_{\nu}|\widetilde{\varphi}_{\beta}\rangle \langle\widetilde{\varphi}_{\sigma}|P_{\nu}|\widetilde{\varphi}_{\gamma}\rangle\nonumber\\
&=\mathpzc{L}^{(\mathrm{nr})}_{~3} + \mathpzc{L}^{(\mathrm{r})}_{~3} 
\label{02}
\end{align}
where 
\begin{equation}
\mathpzc{L}_{~3}^{(\mathrm{r})} =  \sum_{\mathbf{m}} \sum_{(\alpha,\sigma)\in \mathpzc{F}_{\mathbf{m}}} \sum_{(\beta,\gamma)\in \mathpzc{F}_{\mathbf{m}}\big|\begin{smallmatrix}(\beta,\gamma)\neq(\alpha,\sigma) \\(\beta,\gamma)\neq (\sigma,\alpha)\end{smallmatrix}} \langle\widetilde{\varphi}_{\alpha}|P_{\nu}|\widetilde{\varphi}_{\beta}\rangle \langle\widetilde{\varphi}_{\sigma}|P_{\nu}|\widetilde{\varphi}_{\gamma}\rangle,
\label{C1}
\end{equation}
and in Eq.~(\ref{line3}) we have used the identity
\begin{align}
\overline{e^{-i\tau (G_{\mathbf{k}} - G_{\mathbf{l}})}} &= \delta_{G_{\mathbf{k}}G_{\mathbf{l}}}.
\end{align}
Noting that we allow resonance in the system, we do not necessarily have $\delta_{G_{\mathbf{k}}G_{\mathbf{l}}}= \delta_{\mathbf{k}\mathbf{l}}$.

We now rewrite $\mathpzc{L}$ [Eq.~(\ref{eq:L})] using Eqs.~(\ref{degeneracy1}), (\ref{res-reduced-2}), and (\ref{C1}) as
\begin{align}
\mathpzc{L}& = \sum_{\alpha \neq \beta=1}^{D_{\mathrm{E}}} |\langle \widetilde{\varphi}_{\alpha}|P_{\nu}|\widetilde{\varphi}_{\beta} \rangle|^2 + \left(\sum_{\alpha=1}^{D_{\mathrm{E}}} \langle \widetilde{\varphi}_{\alpha}|P_{\nu}|\widetilde{\varphi}_{\alpha} \rangle -\frac{d_{\nu}}{D}\right)^2+ \mathpzc{L}_{~3}^{(\mathrm{r})}\label{eq:X0}\\
&\leqslant \max_{\alpha\neq \beta\in\{1,\ldots,D_{\mathrm{E}}\}} |\langle \varphi_{\alpha}|P_{\nu}|\varphi_{\beta} \rangle|^2 + \max_{\alpha\in\{1,\ldots,D_{\mathrm{E}}\}} \left( \langle \varphi_{\alpha}|P_{\nu}|\varphi_{\alpha} \rangle -\frac{d_{\nu}}{D}\right)^2 +\mathpzc{L}_{~3}^{(\mathrm{r})},
\label{eq:X}
\end{align}
whose RHS is the very quantity $X$ introduced in Eq.~(\ref{eq.}) in the strategy of the proof. We note that the expression for $\mathpzc{L}$ here differs from that in the non-degenerate--non-resonance case \cite{Goldstein-et-al} in that here we have $D_{\mathrm{E}}$ rather than $D$, and a new term $\mathpzc{L}_{~3}^{(\mathrm{r})}$ has emerged. Equation~(\ref{eq:X}) yields 
\begin{align}
\mathbb{E}_{P_{\nu}}[\mathpzc{L}]\leqslant \mathbb{E}_{P_{\nu}}\Big[\max_{\alpha\neq \beta\in\{1,\ldots,D_{\mathrm{E}}\}} |\langle \varphi_{\alpha}|P_{\nu}|\varphi_{\beta} \rangle|^2\Big] + \mathbb{E}_{P_{\nu}}\Big[\max_{\alpha\in\{1,\ldots,D_{\mathrm{E}}\}} \left( \langle \varphi_{\alpha}|P_{\nu}|\varphi_{\alpha} \rangle -\frac{d_{\nu}}{D}\right)^2\Big] + \mathbb{E}_{P_{\nu}}\big[\mathpzc{L}_{~3}^{(\mathrm{r})}\big].
\label{eq:ave-bound}
\end{align}

Following Ref.~\cite{Goldstein-et-al}, the first two terms here can be bounded by using Eqs.~(\ref{eq:l1-1}) and (\ref{eq:l1-2}) of Lemma \ref{lemma:vonneumann1}. In order to set  the scene to employ this lemma, we define the following orthonormal basis set for $\mathpzc{H}$:
\begin{align}
|\Phi_{\kappa}\rangle=\begin{cases}|\varphi_{\kappa}\rangle;~~~&1\leqslant\kappa\leqslant D_{\mathrm{E}}\\
|\varphi^{\perp}_{\kappa}\rangle;~~~&D_{\mathrm{E}}+1\leqslant\kappa\leqslant D,
\end{cases}
\label{def:Phi}
\end{align}
where $\{|\varphi^{\perp}_{\kappa}\rangle\}$ are some vectors orthogonal to $\{|\varphi_{\alpha}\rangle\}$, chosen to complete the basis set, and thus $\langle \Phi_{\kappa}|\Phi_{\iota}\rangle=\delta_{\kappa \iota}$. It is evident from Eq.~(\ref{eq:ave-bound}) that
\begin{align}
\mathbb{E}_{P_{\nu}}[\mathpzc{L}]\leqslant \mathbb{E}_{P_{\nu}}\Big[\max_{\kappa\neq \iota\in\{1,\ldots,D\}} |\langle \Phi_{\kappa}|P_{\nu}|\Phi_{\iota} \rangle|^2\Big] + \mathbb{E}_{P_{\nu}}\Big[\max_{\kappa\in\{1,\ldots,D\}} \left( \langle \Phi_{\kappa}|P_{\nu}|\Phi_{\kappa} \rangle -\frac{d_{\nu}}{D}\right)^2\Big] + \mathbb{E}_{P_{\nu}}\big[\mathpzc{L}_{~3}^{(\mathrm{r})}\big].
\label{eq:ave-bound-2}
\end{align} 
Now we choose another orthonormal basis set for $\mathpzc{H}$ as $\{|\omega_{\iota}\rangle\}_{\iota=1}^{D} $ such that $P_{\nu}=\sum_{\iota\in \mathpzc{J}_{\nu}}|\omega_{\iota}\rangle\langle \omega_{\iota}|$, where $\mathpzc{J}_{\nu}$ is the set of indices associated with the spectral representation of $P_{\nu}$, with $|\mathpzc{J}_{\nu}|=d_{\nu}$, for all $\nu\in\{1,\ldots,M\}$. Hence if we define the unitary matrix $U_{\kappa \iota}=\langle \Phi_{\kappa}|\omega_{\iota}\rangle$, we can use Lemma \ref{lemma:vonneumann1} to obtain
\begin{align}
\mathbb{E}_{P_{\nu}}\Big[\max_{\alpha\neq \beta\in\{1,\ldots,D_{\mathrm{E}}\}} |\langle \varphi_{\alpha}|P_{\nu}|\varphi_{\beta} \rangle|^2\Big] + \mathbb{E}_{P_{\nu}}\Big[\max_{\alpha\in\{1,\ldots,D_{\mathrm{E}}\}} \left( \langle \varphi_{\alpha}|P_{\nu}|\varphi_{\alpha} \rangle -\frac{d_{\nu}}{D}\right)^2\Big]&\leqslant \frac{\log D}{D} + \frac{9d_{\nu}\log D}{D^2}\nonumber\\
&\overset{\mathrm{Eq.~(\ref{condition-c})}}{<}\frac{10\log D}{D}.
\label{logD}
\end{align}

Now we show how to derive a bound on $\mathpzc{L}_{~3}^{(\mathrm{r})}$ and $\mathbb{E}_{P_{\nu}}[\mathpzc{L}_{~3}^{(\mathrm{r})}]$ [in Eq.~(\ref{eq:ave-bound})]. From positivity of $P_{\nu}$ (and hence $P_{\nu} \otimes P_{\nu}$), 
\begin{equation}
(\langle\widetilde{\varphi}_{\alpha}|\otimes\langle\widetilde{\varphi}_{\sigma}|- \langle\widetilde{\varphi}_{\beta}|\otimes\langle\widetilde{\varphi}_{\gamma}|)P_{\nu}\otimes P_{\nu}(|\widetilde{\varphi}_{\alpha}\rangle\otimes|\widetilde{\varphi}_{\sigma}\rangle - |\widetilde{\varphi}_{\beta}\rangle\otimes |\widetilde{\varphi}_{\gamma}\rangle) \geqslant 0,
\label{C2}
\end{equation}
we obtain
\begin{equation}
\mathrm{Re}\big[\langle\widetilde{\varphi}_{\alpha}|P_{\nu}|\widetilde{\varphi}_{\beta}\rangle \langle\widetilde{\varphi}_{\sigma}|P_{\nu}|\widetilde{\varphi}_{\gamma}\rangle\big] \leqslant \frac{1}{2} \big( \langle \widetilde{\varphi}_{\alpha}|P_{\nu}|\widetilde{\varphi}_{\alpha}\rangle \langle \widetilde{\varphi}_{\sigma}|P_{\nu}|\widetilde{\varphi}_{\sigma}\rangle + \langle \widetilde{\varphi}_{\beta}|P_{\nu}|\widetilde{\varphi}_{\beta}\rangle \langle \widetilde{\varphi}_{\gamma}|P_{\nu}|\widetilde{\varphi}_{\gamma}\rangle \big).
\label{C3}
\end{equation}
Combining Eqs.~(\ref{C1}) and (\ref{C3}) yields
\begin{equation}
\mathpzc{L}_{~3}^{(\mathrm{r})} \leqslant  \frac{1}{2} \sum_{\mathbf{m}} \sum_{(\alpha,\sigma)\in \mathpzc{F}_{\mathbf{m}}} \sum_{(\beta,\gamma)\in \mathpzc{F}_{\mathbf{m}}\big|\begin{smallmatrix}(\beta,\gamma)\neq(\alpha,\sigma) \\(\beta,\gamma)\neq (\sigma,\alpha)\end{smallmatrix}} \big( \langle \widetilde{\varphi}_{\alpha}|P_{\nu}|\widetilde{\varphi}_{\alpha}\rangle \langle \widetilde{\varphi}_{\sigma}|P_{\nu}|\widetilde{\varphi}_{\sigma}\rangle + \langle \widetilde{\varphi}_{\beta}|P_{\nu}|\widetilde{\varphi}_{\beta}\rangle \langle \widetilde{\varphi}_{\gamma}|P_{\nu}|\widetilde{\varphi}_{\gamma}\rangle \big).
\label{C4}
\end{equation}
For every pair $(\alpha,\sigma)\in \mathpzc{F}_{\mathbf{m}} $, the pair $(\beta,\gamma)$ can have $(f_{\mathbf{m}}-2)$ different values. Furthermore, we have obtained two terms on the RHS of Eq.~(\ref{C4}) which are equal. Thus
\begin{align}
\mathpzc{L}_{~3}^{(\mathrm{r})} &\leqslant \sum_{\mathbf{m}} [(f_{\mathbf{m}}-2)] \sum_{(\alpha,\sigma)\in \mathpzc{F}_{\mathbf{m}}} \langle \widetilde{\varphi}_{\alpha}|P_{\nu}|\widetilde{\varphi}_{\alpha}\rangle \langle \widetilde{\varphi}_{\sigma}|P_{\nu}|\widetilde{\varphi}_{\sigma}\rangle
\nonumber
\\
&\leqslant (D_{\mathrm{F}}-2)\sum_{\mathbf{m}} \big(\sum_{(\alpha,\sigma)\in \mathpzc{F}_{\mathbf{m}}} \langle \widetilde{\varphi}_{\alpha}|P_{\nu}|\widetilde{\varphi}_{\alpha}\rangle \langle \widetilde{\varphi}_{\sigma}|P_{\nu}|\widetilde{\varphi}_{\sigma}\rangle\big)
\nonumber\\
&= (D_{\mathrm{F}}-2)\sum_{\alpha\sigma} \langle \widetilde{\varphi}_{\alpha}|P_{\nu}|\widetilde{\varphi}_{\alpha}\rangle \langle \widetilde{\varphi}_{\sigma}|P_{\nu}|\widetilde{\varphi}_{\sigma}\rangle
\nonumber\\
&= (D_{\mathrm{F}}-2)\sum_{\alpha\sigma} |c_{\alpha}|^2 |c_{\sigma}|^2 \langle \varphi_{\alpha}|P_{\nu}|\varphi_{\alpha}\rangle \langle \varphi_{\sigma}|P_{\nu}|\varphi_{\sigma}\rangle.
\label{C8}
\end{align}
To get the third inequality we used the fact that each pair $(\alpha,\sigma)$ appears only in one $\mathbf{m}$-shell. It should be noted that the $\alpha = \sigma$ case has automatically been included in our calculations, thus we do not need to consider it separately. Equation~(\ref{C8}) gives
\begin{align}
\mathbb{E}_{P_{\nu}}\big[\mathpzc{L}_{~3}^{(\mathrm{r})} \big] &\leqslant  (D_{\mathrm{F}}-2) ~\mathbb{E}_{P_{\nu}}\Big[ \sum_{\alpha\sigma}  |c_{\alpha}|^2  |c_{\sigma}|^2  \langle \varphi_{\alpha}|P_{\nu}|\varphi_{\alpha}\rangle \langle \varphi_{\sigma}|P_{\nu}|\varphi_{\sigma}\rangle \Big]
\nonumber\\
&= (D_{\mathrm{F}}-2) \sum_{\alpha\sigma}  |c_{\alpha}|^2  |c_{\sigma}|^2~ \mathbb{E}_{P_{\nu}}\big[ \langle \varphi_{\alpha}|P_{\nu}|\varphi_{\alpha}\rangle \langle \varphi_{\sigma}|P_{\nu}|\varphi_{\sigma}\rangle \big].
\label{G44}
\end{align}
Since $P_{\nu}$ has a \textit{fixed} rank, averaging over random $P_{\nu}$ can be equivalently replaced with averaging over random unitary operators $U\in\mathrm{SU}(D)$ according to the associated Haar measure, acting on a fixed projection $P_{\nu_0}$
\begin{align}
\mathbb{E}_{P_{\nu}}\big[\mathpzc{L}_{~3}^{(\mathrm{r})}\big] & = \mathbb{E}_{U}\big[\mathpzc{L}_{~3}^{(\mathrm{r})}|_{P_{\nu_0}}\big]\nonumber\\
 &\leqslant (D_{\mathrm{F}}-2) \sum_{\alpha\sigma}  |c_{\alpha}|^2  |c_{\sigma}|^2 ~\mathbb{E}_{U}\big[ \langle \varphi_{\alpha}|U^{\dagger}P_{\nu_0}U|\varphi_{\alpha}\rangle \langle \varphi_{\sigma}|U^{\dagger}P_{\nu_0} U|\varphi_{\sigma}\rangle \big].
\label{G45}
\end{align}
Since $U$ is chosen based on the Haar measure of $\mathrm{SU}(D)$, thus by applying random unitaries on $|\varphi_{\alpha}\rangle$ (a specific/fixed vector in $P_{\nu}\mathpzc{H}$) we indeed obtain vectors $|\xi\rangle$ which are uniformly-randomly distributed all over the $D$-dimensional Hilbert space $\mathpzc{H}$ \cite{ref-Muller}. Hence we can rewrite Eq.~(\ref{G45}) as 
\begin{align}
\mathbb{E}_{P_{\nu}}\big[\mathpzc{L}_{~3}^{(\mathrm{r})}\big] &\leqslant (D_{\mathrm{F}}-2) \sum_{\alpha\gamma}  |c_{\alpha}|^2  |c_{\gamma}|^2 ~\mathbb{E}_{\xi}\big[ \Vert P_{\nu_0}|\xi_1\rangle\Vert^2 \Vert P_{\nu_0} |\xi_{2}\rangle\Vert^2 \big]\nonumber\\
&\overset{\mathrm{Eq.~(\ref{eq:e-v})}}{\leqslant}  (D_{\mathrm{F}}-2) \left(\mathbb{E}_{\xi}\big[ \Vert P_{\nu_0}|\xi\rangle\Vert^4\big] +\mathbb{V}_{\xi}\big[\Vert P_{\nu_0}|\xi\rangle\Vert^2 \big]\right)\nonumber\\
&\overset{\mathrm{Lemma~\ref{lemma:vonneumann2}}}{\leqslant} (D_{\mathrm{F}}-2)\Big(\frac{d_{\nu}}{D}\Big) \Big( \frac{d_{\nu}+1}{D+1}\Big) \nonumber\\
&\overset{1\ll d_{\nu}\ll D}{\lessapprox} (D_{\mathrm{F}}-2)\Big(\frac{d_{\nu}}{D}\Big)^2.
\label{eq:boundl3}
\end{align}
It is evident that $\mathbb{E}_{P_{\nu}}[\mathpzc{L}_{~3}^{(\mathrm{r})}]$ becomes negligible if
\begin{equation}
(D_{\mathrm{F}}-2) \Big(\frac{d_{\nu}}{D}\Big)^2 \ll 1.
\label{G411}
\end{equation}

We insert Eqs.~(\ref{logD}) and (\ref{eq:boundl3}) into Eq.~(\ref{eq:ave-bound}), which yields
\begin{align}
\mathbb{E}_{P_{\nu}}[\mathpzc{L}]\leqslant \frac{10\log D}{D} + (D_{\mathrm{F}}-2)\Big(\frac{d_{\nu}}{D}\Big)^{2},
\label{eq:ave-bound-final}
\end{align}
whose RHS is the very $\mathbb{E}_{P_{\nu}}[X]$ introduced in the beginning of this subsection in the strategy of the proof. Next, following Eq.~(\ref{MM}), we require that the RHS of Eq.~(\ref{eq:ave-bound-final}) to be upperbounded by $B\delta$; that is, 
\begin{align}
\frac{M^2 D}{\delta' \varepsilon^2 d_{\nu}}\Big[\frac{10\log D}{D} + (D_{\mathrm{F}}-2)\Big(\frac{d_{\nu}}{D}\Big)^{2}\Big] < \delta,
\end{align}
or equivalently
\begin{align}
\frac{M^2}{\delta \delta' \varepsilon^2} \Big[\frac{10\log D}{D} + (D_{\mathrm{F}}-2) \Big( \frac{d_{\nu}}{D}\Big)^2  \Big] < \frac{d_{\nu}}{D}.
\label{eq:condition-d-1}
\end{align}
On the other hand, in order to use Lemma \ref{lemma:vonneumann1}---in calculating the averages---we require condition (\ref{condition-c}). If we combine these conditions, we obtain
\begin{align}
\max\Big\{C,\frac{10M^2}{\delta \delta' \varepsilon^2} \Big[1+ \frac{(D_{\mathrm{F}}-2)d_{\nu}^2}{10D\log D} \Big] \Big\}\frac{\log D}{D}< \frac{d_{\nu}}{D} <\frac{1}{C}.
\label{condition-final}
\end{align}

\hfill $\blacksquare$

\section{Summary}
\label{label:conc}

It has been known that for the validity of the original quantum ergodic theorem, non-degeneracy and non-resonance properties for their Hamiltonians are among the conditions, which hold in typical systems. Here we, however, have considered systems which lack these conditions. We have proved a modified version of the quantum ergodic theorem which concerns validity of the ergodic hypothesis and normal typicality in atypical systems. We have shown that degeneracy does not considerably modify the condition of normal typicality or ergodicity, whereas the existence of resonance is more dominant for obstructing ergodicity. The effect of the non-resonance condition has been shown to come through degeneracy of energy sum structure of the system of question.   

\begin{acknowledgements}
P.A. acknowledges useful discussions with P. Delgosha.
\end{acknowledgements}

\appendix 

\section{Proof that $D_{\mathrm{F}}=2$ for the non-resonant case}
\label{app:df}

The non-resonance condition implies that (see subsection \ref{subsec:gap} or Fig.~\ref{fig:gap})
\begin{equation}
g_{\mathbf{k}}\equiv |\mathpzc{G}_{\mathbf{k}}|=1,~~~\forall \mathbf{k}\neq\mathbf{0},
\label{eq:nrc-g}
\end{equation}
whence $D_{\mathrm{G}}=1$. For example, if for some $\alpha$, $\beta$, $\gamma$, and $\sigma$ we have 
\begin{equation}
E_{\alpha} - E_{\gamma} = E_{\beta} - E_{\sigma} \equiv G_{\mathbf{k}},
\label{eq:nr1}
\end{equation}
for some $\mathbf{k}$ (or equivalently $E_{\alpha} -E_{\beta} = E_{\gamma} - E_{\sigma} \equiv G_{\mathbf{l}}$), then
\begin{equation}
(\alpha,\gamma) = (\beta,\sigma)~~~\mathrm{or}~~~(\alpha,\beta)=(\gamma,\sigma),
\label{eq:nr2}
\end{equation}
or equivalently
\begin{align}
\mathpzc{G}_{\mathbf{k}} &=\{(\alpha,\gamma)\},~~~g_{\mathbf{k}}=1,\\
\mathpzc{G}_{\mathbf{l}} &=\{(\alpha,\beta)\},~~~g_{\mathbf{l}}=1.
\end{align}
Equation (\ref{eq:nr1}) can be recast in the following form too:
\begin{equation}
E_{\alpha} + E_{\sigma} = E_{\beta} + E_{\gamma} \equiv F_{\mathbf{m}},
\end{equation}
for some $\mathbf{m}$, which means
\begin{equation}
\{(\alpha,\sigma),(\sigma,\alpha),(\beta,\gamma),(\gamma,\beta)\}\in\mathpzc{F}_{\mathbf{m}}.
\label{eq:nr3}
\end{equation}
But the relations in Eq.~(\ref{eq:nr2}) both imply that 
\begin{equation}
(\alpha,\sigma)=(\beta,\gamma).
\end{equation}
Note that there cannot be any other pair in $\mathpzc{F}_{\mathbf{m}}$ except those listed in Eq.~(\ref{eq:nr3}). Hence the non-resonance condition (\ref{eq:nrc-g}) can be equivalently rewritten as 
\begin{equation}
f_{\mathbf{m}}\equiv |\mathpzc{F}_{\mathbf{m}}|=2,~~~\forall \mathbf{m},
\label{eq:nrc-f}
\end{equation}
which in turn implies $D_{\mathrm{F}}=2$. 

\textit{Remark.---}Note that the existence of degeneracy contributes in our results in the paper mainly through modifying the estimates on the resulting resonance effects so that such effects become dominant.

\section{Calculating the average and variance of $\Vert P_{\nu}|\varphi_{\alpha}\rangle\Vert^2$ with uniform measure on $P_\nu $}
\label{app:averages}

Here we reproduce the average and variance of $\Vert P_{\nu}|\varphi_{\alpha}\rangle\Vert^2$, following the approach of Ref.~\cite{book:Jancel}. Let us take $\{|\omega_{\iota}\rangle\}_{i=1}^{D}$ as an orthonormal basis for the $D$-dimensional Hilbert space $\mathpzc{H}$ such that  
\begin{equation}
P_\nu =\sum_{\iota\in\mathpzc{J}_{\nu}}|\omega_{\iota}\rangle \langle \omega_{\iota}|,
\label{Pnu}
\end{equation}
where $\mathpzc{J}_{\nu}\subset\{1,\ldots,D\}$ constitutes the set of indices associated with $P_{\nu}$, and $|\mathpzc{J}_{\nu}|=d_{\nu}$. We introduce another orthonormal basis $\{|\Phi_{\kappa}\}_{\iota=1}^{D}$ for $\mathpzc{H}$ related to $\{|\varphi_{\alpha}\rangle \}_{\alpha=1}^{D_{\mathrm{E}}}$ as in Eq.~(\ref{def:Phi}). Now let us expand $|\omega_{\iota}\rangle$ as 
\begin{equation}
|\omega_{\iota} \rangle=\sum_{\kappa=1}^{D} (x_{\kappa}+iy_{\kappa}) |\Phi_{\kappa}\rangle,
\label{omega}
\end{equation} 
where the normalization condition $\langle \omega_{\iota}|\omega_{\iota}\rangle=1$ reads as $\sum_{\kappa=1}^{D} (x_{\kappa}^2+y_{\kappa}^2)=1$. This is the equation of the $2D$-dimensional unit hypersphere $S^{2D}$. In addition, we have
\begin{equation}
\Vert P_{\nu}|\varphi_{\alpha}\rangle\Vert^2 = \sum_{\kappa\in\mathpzc{J}_{\nu}}(x_{\kappa}^{2}+y_{\kappa}^2).
\end{equation}
Thus averaging over fixed-rank $P_{\nu}$s with uniform distribution is equivalent to averaging uniformly over $S^{2D}$. This yields
\begin{align}
\mathbb{E}_{P_{\nu}}[\Vert P_{\nu}|\varphi_{\alpha}\rangle\Vert^2]
&=2|\mathpzc{J}_{\nu}|~ \mathbb{E}_{S^{2D}}[x_{\kappa}^2],
\label{app ave4}
\end{align}
where $\mathbb{E}_{S^{2D}}[Y]\equiv[1/A_{2D}]\oint_{S^{2D}}\mathrm{d}\sigma~Y$ indicates the average over $S^{2D}$, in which $\mathrm{d}\sigma$ and $A_{2D}$ are the surface element and the total area of the unit $2D$-dimensional hypersphere. Straightforward calculations yield \cite{book:Jancel}
\begin{align}
\mathbb{E}_{S^{2D}}[x_{\kappa}^2]&=\frac{1}{2D},
\label{app ave2}\\
\mathbb{V}_{S^{2D}}[x_{\kappa}^2]&=\frac{(D-1)}{D^2(D+1)},
\label{app var2} \\
\mathbb{C}_{S^{2D}}[x_{\kappa}^2,x_{\iota}^2]&=-\frac{1}{D^2(D+1)},~~\kappa\neq\iota,
\label{app cov}
\end{align}
whence
\begin{align}
\mathbb{E}_{P_{\nu}}\big[\Vert P_{\nu}|\varphi_{\alpha}\rangle\Vert^2\big]&=
\frac{d_{\nu}}{D},
\label{app ave3}\\
\mathbb{V}_{P_{\nu}}[\Vert P_{\nu}|\varphi_{\alpha}\rangle\Vert^2]&\overset{\mathrm{Eq.~(\ref{var-covar})}}{=}
\frac{1}{d_{\nu}}\left(\frac{d_{\nu}}{D}\right)^2 \frac{(D-d_{\nu})}{(D+1)}.
\label{app var3}
\end{align}



\end{document}